\documentclass[12pt]{article}
\usepackage{amssymb}
\usepackage{color,graphicx}
\usepackage{amsmath}

\newcommand{\ep}{\varepsilon}
\newcommand{\bea}{\begin{eqnarray}}
\newcommand{\eea}{\end{eqnarray}}
\newcommand{\be}{\begin{equation}}
\newcommand{\ee}{\end{equation}}

\begin{document}

\begin{center}
{\Large {\bf The property of maximal transcendentality:
calculation of 
master integrals
}}
\\ \vspace*{5mm} A.~V.~Kotikov
\end{center}

\begin{center}
Bogoliubov Laboratory of Theoretical Physics \\
Joint Institute for Nuclear Research\\
141980 Dubna, Russia
\end{center}

\begin{abstract}
We review 
the results 
having the property of maximal transcendentality.

\end{abstract}


\section{Introduction}


Recently discovered that a popular  property of maximal transcendentality, which was introduced in \cite{KL}
for the Balitsky-Fadin-Kuraev-Lipatov (BFKL) kernel \cite{BFKL,next} in the  ${\mathcal N}=4$
Supersymmetric Yang-Mills (SYM) model \cite{BSSGSO}, is also applicable to the amplitudes,
form-factors and correlation functions (see \cite{Eden:2011we,Schlotterer:2012ny,Eden:2012rr}
and discussions and references therein).

The aim
of this short paper is 
to show this property in the results for the anomalous dimension (AD) matrix
of the twist-2 Wilson operators and to demonstrate a similar feature in  the results for 
so-called  master integrals \cite{Broadhurst:1987ei}.

\section{ADs
in ${\mathcal N}=4$ SYM}

The ADs
govern the
Bjorken scaling violation for parton distributions ($\equiv$ matrix elemens of the twist-2 Wilson operators)
in a framework of Quantum Chromodynamics (QCD).

\begin{figure}[t]
\includegraphics[width=0.3\textwidth,height=0.6\textheight,angle=92]{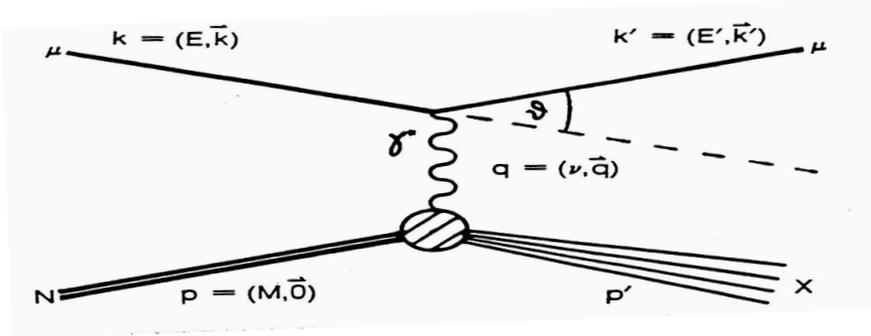}
\caption{
 The 
deep inelastic muon-nucleon scattering, where $k$, $q$ and $p$ are the muon,
photon and nucleon momenta, respectively. In the  deep inelastic
kinematics, 
$p^2=M^2 \to 0$, where $M$ is the nucleon mass. The standard variables 
are $Q^2=-q^2>0$ and the Bjorken variable $x=Q^2/(2pq)$, where $Q^2$ is the 
``mass'' of the virtual photon and $x$ is the part of the nucleon momentum
carried by the colliding parton (quark or gluon).
}
\end{figure}

The BFKL
and Dokshitzer-Gribov-Lipatov-Altarelli-Parisi (DGLAP) \cite{DGLAP}
DGLAP equations resum, respectively, the most important 
contributions
$\sim \alpha_s \ln(1/x)$ and $\sim \alpha_s \ln(Q^2/\Lambda^2)$ in different
kinematical regions of the Bjorken variable $x$ and the ``mass'' $Q^2$ of the
virtual photon in the 
deep inelastic lepton-hadron scattering  (see Fig. 1 for the 
muon-nucleon case) 
and, thus, they are the cornerstone in analyses of
the experimental data
from lepton-nucleon and nucleon-nucleon scattering processes.
In the supersymmetric generalization of QCD
the equations are simplified drastically (see \cite{KL00}).


\subsection{Leading order 
}

The elements of the leading order (LO) AD
matrix in the ${\mathcal N}=4$
SYM have
the following form (see \cite{LN4}):
\begin{eqnarray}
\gamma^{(0)}_{gg}(j) &=& 4
\left( \Psi(1)-\Psi(j-1)-\frac{2}{j}+\frac{1}{j+1}
-\frac{1}{j+2} \right),  \nonumber \\
\gamma^{(0)}_{\lambda g}(j) &=& 8 \left(\frac{1}{j}-\frac{2}{j+1}+\frac{2}{j+2}
\right),~~~~~~~~~~\, \gamma^{(0)}_{\varphi g}(j) ~=~ 12 \left( \frac{1}{j+1}
-\frac{1}{j+2} \right),  \nonumber \\
\gamma^{(0)}_{g\lambda}(j) &=& 2 \left(\frac{2}{j-1}-\frac{2}{j}+\frac{1}{j+1}
\right),~~~~~~~~~~\, \gamma^{(0)}_{q\varphi}(j) ~=~ \frac{8}{j} \,,
\nonumber \\
\gamma^{(0)}_{\lambda \lambda}(j) &=& 4 \left( \Psi(1)-\Psi(j)+\frac{1}{j}-
\frac{2}{j+1}%
\right),~~ \gamma^{(0)}_{\varphi \lambda}(j) ~=~ \frac{6}{j+1} \,, \nonumber \\
\gamma^{(0)}_{\varphi \varphi}(j) &=& 4 \left( \Psi(1)-\Psi(j+1)\right),
~~~~~~~~~~~~~~~\, \gamma^{(0)}_{g\varphi}(j) ~=~ 4
\left(\frac{1}{j-1}-\frac{%
1}{j} \right),  \label{3.2}
\end{eqnarray}
where $j$ is the Mellin moment 
(or spin) number.


The matrix, based on the ADs
(\ref{3.2}),
can be diagonalized \cite{LN4,KL}:
\begin{eqnarray}
{\Biggl[D\Gamma D^{-1}\Biggr]}^{N=4}_{\mathbf{unpol}} =
\begin{array}{|ccc|}
-4S_1(j-2) & 0 & 0 \\
0  & -4S_1(j) & 0  \\
0 & 0 & -4S_1(j+2)
\end{array}
\,, \nonumber
\end{eqnarray}
where $S_1(j)$ is defined below in (\ref{FI2}).

Thus, the LO ADs
of all multiplicatively renormalized Wilson operators
can be extracted through one universal function
$$
\gamma^{(0)}_{uni}(j)~=~-4S(j-2) \equiv
-4\Bigl(\Psi(j-1)-\Psi(1) \Bigr) \equiv -4 \sum_{r=1}^{j-2}
\frac{1}{r}. $$
Same results can be obtained also for spin-dependent case (see \cite{LN4,KL}).

\subsection{Method to get the universal AD
}\label{MethodAD}

Let us to introduce the
transcendentality level $i$ for the 
harmonic sums
\be
S_{\pm a}(j)\ =\ \sum^j_{m=1} \frac{(\pm 1)^m}{m^a},
\ \ S_{\pm a,\pm b,\pm c,\cdots}(j)~=~ \sum^j_{m=1}
\frac{(\pm 1)^m}{m^a}\, S_{\pm b,\pm c,\cdots}(m),  \label{FI2}
\ee
and Euler-Zagier constants
\be
\zeta(\pm a)\ =\ \sum^{\infty}_{m=1} \frac{(\pm 1)^m}{m^a},
\ \ \zeta(\pm a,\pm b,\pm c,\cdots )~=~ \sum^{\infty}_{m=1}
\frac{(\pm 1)^m}{m^a}\, S_{\pm b,\pm c,\cdots}(m-1),  \label{Euler}
\ee
in the following way
\be
S_{\pm a,\pm b,\pm c,\cdots}(j) \sim \zeta(\pm a,\pm b,\pm c,\cdots )
\sim 1/j^i, ~~~~(i=a+b+c+ \cdots)  \label{Tran}
\ee

Then, the basic functions $\gamma
_{uni}^{(0)}(j)$, $\gamma _{uni}^{(1)}(j)$ and $\gamma _{uni}^{(2)}(j)$ are
assumed to be of the types $\sim 1/j^{i}$ with the levels $i=1$, $i=3$ and
$i=5$, respectively. An exception could be for the terms appearing at a given
order from previous orders of the perturbation theory. Such
contributions could be generated and/or removed by an approximate finite
renormalization of the coupling constant. But these terms do not appear in
the ${\overline{\mathrm{DR}}}$-scheme \cite{DRED}.

It is known, that at the LO, the next-to-leading order (NLO) and the 
next-to-next-to-leading order (NNLO) approximations
(with the SUSY relation for the QCD color factors $C_{F}=C_{A}=N_{c}$) the
most complicated contributions (with $i=1,~3$ and $5$, respectively) are the
same for all LO, NLO and NNLO ADs
in QCD~\cite{VMV}
and for the LO and NLO scalar-scalar Ads
\cite{KoLiVe}. This property allows one to find the
universal ADs
$\gamma _{uni}^{(0)}(j)$, $\gamma _{uni}^{(1)}(j)$ and 
$\gamma_{uni}^{(2)}(j)$ without knowing all elements of the AD
matrix~\cite{KL}, which was verified for $\gamma _{uni}^{(1)}(j)$ by the exact 
calculations in~\cite{KoLiVe}.

\subsection{Universal AD
for ${\mathcal N}=4$ SYM}

The final three-loop result
\footnote{
Note, that in an accordance with Ref.~\cite{next}
 our normalization of $\gamma (j)$ contains
the extra factor $-1/2$ in comparison with
the standard normalization (see~\cite{KL})
and differs by sign in comparison with one from Ref.~\cite{VMV}.}
for the universal AD
$\gamma_{uni}(j)$
for ${\mathcal N}=4$ SYM is~\cite{KLOV}
\begin{eqnarray}
\gamma(j)\equiv\gamma_{uni}(j) ~=~ \hat a \gamma^{(0)}_{uni}(j)+\hat a^2
\gamma^{(1)}_{uni}(j) +\hat a^3 \gamma^{(2)}_{uni}(j) + ... , \qquad \hat a=\frac{\alpha N_c}{4\pi}\,,  \label{uni1}
\end{eqnarray}
where
\begin{eqnarray}
\frac{1}{4} \, \gamma^{(0)}_{uni}(j+2) &=& - S_1,  \label{uni1.1} \\
\frac{1}{8} \, \gamma^{(1)}_{uni}(j+2) &=& \Bigl(S_{3} + \overline S_{-3} \Bigr) -
2\,\overline S_{-2,1} + 2\,S_1\Bigl(S_{2} + \overline S_{-2} \Bigr),  \label{uni1.2} \\
\frac{1}{32} \, \gamma^{(2)}_{uni}(j+2) &=& 2\,\overline S_{-3}\,S_2 -S_5 -
2\,\overline S_{-2}\,S_3 - 3\,\overline S_{-5}  +24\,\overline S_{-2,1,1,1}\nonumber\\
&&\hspace{-1.5cm}+ 6\biggl(\overline S_{-4,1} + \overline S_{-3,2} + \overline S_{-2,3}\biggr)
- 12\biggl(\overline S_{-3,1,1} + \overline S_{-2,1,2} + \overline S_{-2,2,1}\biggr)\nonumber \\
&& \hspace{-1.5cm}  -
\biggl(S_2 + 2\,S_1^2\biggr) \biggl( 3 \,\overline S_{-3} + S_3 - 2\, \overline S_{-2,1}\biggr)
- S_1\biggl(8\,\overline S_{-4} + \overline S_{-2}^2\nonumber \\
&& \hspace{-1.5cm}  + 4\,S_2\,\overline S_{-2} +
2\,S_2^2 + 3\,S_4 - 12\, \overline S_{-3,1} - 10\, \overline S_{-2,2} + 16\, \overline S_{-2,1,1}\biggr)
\label{uni1.5}
\end{eqnarray}
with $S_{\pm a,\pm b, \pm c,...} \equiv  S_{\pm a,\pm b,\pm c,...}(j)$ and 
\begin{eqnarray}
\overline S_{-a,b,c,\cdots}(j) ~=~ (-1)^j \, S_{-a,b,c,...}(j)
+ S_{-a,b,c,\cdots}(\infty) \, \Bigl( 1-(-1)^j \Bigr).  \label{ha3}
\end{eqnarray}

The expression~(\ref{ha3}) is the analytical continuation (to real and 
complex $j$) \cite{KK} of the harmonic sums $S_{-a,b,c,\cdots}(j)$.

The results for $\gamma^{(3)}_{uni}(j)$ \cite{KLRSV,KoReZi} and $\gamma^{(4)}_{uni}(j)$ \cite{LuReVe}
can be obtained from the long-range asymptotic Bethe equations \cite{Staudacher:2004tk} 
for twist-two operators and the additional 
contribution
of the wrapping
corrections.


\section{Calculation of Feynman integrals}

The arguments similar to ones considered in \cite{KL}, 
give a possibility to calculate
a large class of Feynman
diagrams,
so-called master-integrals \cite{Broadhurst:1987ei}.
Let us consider it in some details.

Application of the integration-by-part (IBP) procedure \cite{Chetyrkin:1981qh}
to loop internal momenta
leads to
relations between different Feynman integrals (FIs) and, thus,
to necessity to calculate only some of them, which in a sense, are
independent (see  \cite{Kotikov:1990zs}). These independent diagrams
(which were chosen quite arbitrary, of course) are called the
master-integrals \cite{Broadhurst:1987ei}.

The application of the IBP procedure \cite{Chetyrkin:1981qh} to the
master-integrals themselves leads to the differential equations
\cite{Kotikov:1990zs,Kotikov:1990kg} for them with
the inhomogeneous terms (ITs) containing less complicated diagrams.
\footnote{The ``less complicated diagrams'' contain usually less number of
propagators and sometimes they can be represented as diagrams with less
number of loops and with some ``effective masses'' (see, for example, 
\cite{Kniehl:2005bc} and references therein).} The application
of the IBP procedure to the diagrams for ITs leads to the new differential
equations for them with the new ITs
containing even farther
less complicated diagrams.
Repeating the procedure several times, at a last step one can obtain the
ITs containing mostly
tadpoles which can be calculated  in-turn very easily.

Solving the
differential equations
at this last step,
one can reproduce the diagrams for ITs of the differential equations at the
previous step.
Repeating the procedure several times one can obtain the results for the
initial FIs.

This scheme has been used successfully for calculation of two-loop two-point
\cite{Kotikov:1990zs,Kotikov:1990kg,Fleischer:1999hp}
and three-point diagrams \cite{Fleischer:1997bw,FleKoVe}
with one nonzero mass. This procedure is very
powerful but quite complicated. There are, however, some simplifications, which
are based on the series representations of FIs.

\begin{figure}[t]
\includegraphics[width=0.6\textwidth,height=0.3\textheight,angle=0]{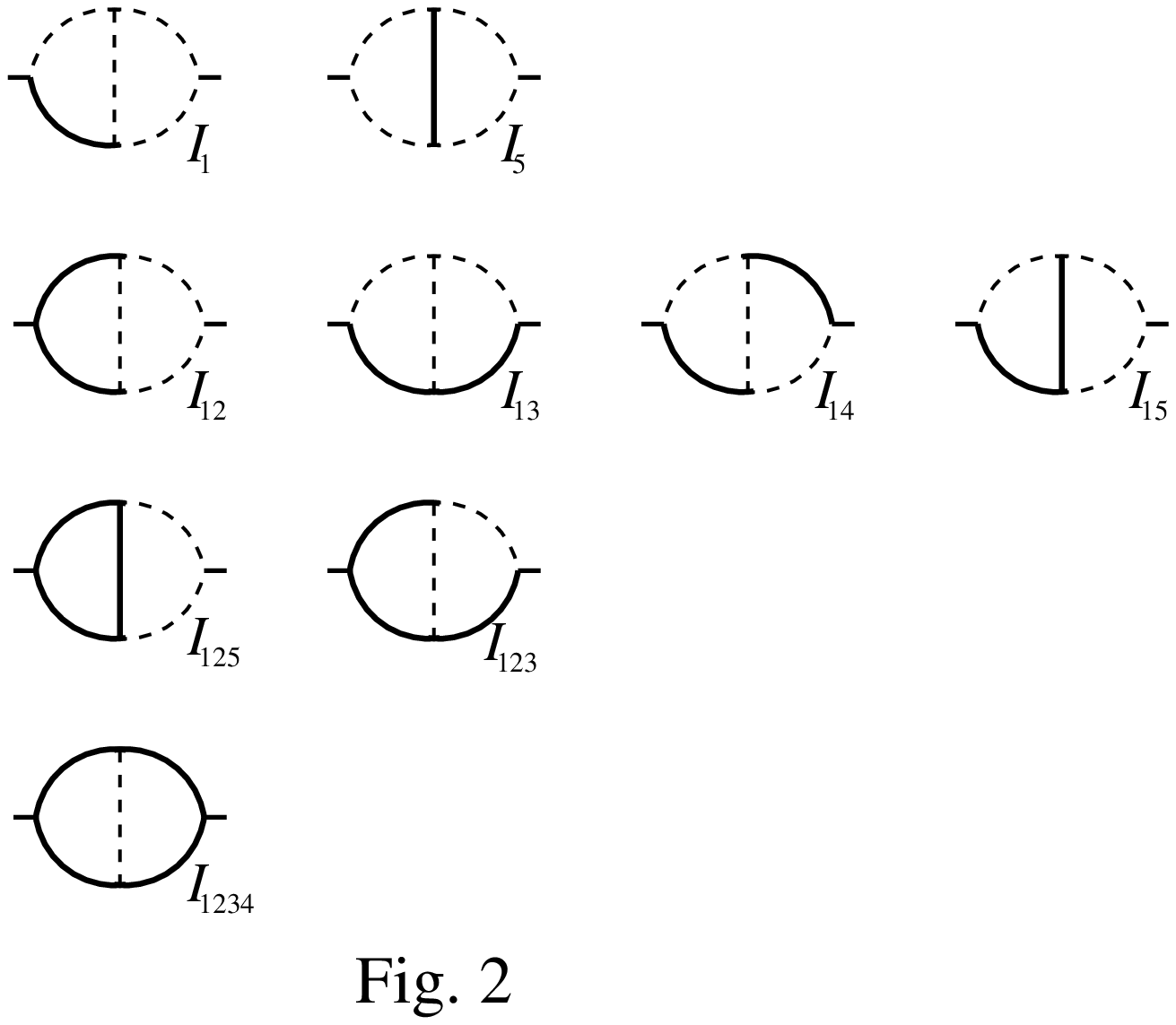}
\end{figure}

Indeed, the inverse-mass expansion of two-loop two-point (see Fig. 2)
and three-point
diagrams (see Fig. 3)
\footnote{We consider only three-point
diagrams with independent upward momenta $q_1$ and $q_2$, which obey
the conditions
 $q_1^2=q_2^2=0$ and $(q_1+q_2)^2\equiv q^2 \neq 0$, where $q$ is downward
momentum.}
with one nonzero mass (massless and massive propagators are shown as dashed and solid lines,
respectively),
can be considered as
\begin{eqnarray}
&&\mbox{ FI} ~ 
= ~ \frac{\hat{N}}{q^{2\alpha}} \,
\sum_{n=1} \, C_n \, \frac{{(\eta x)}^n}{n^c} \, \biggl\{F_0(n) +
\biggl[ \ln (-x) \, F_{1,1}(n)  +
\frac{1}{\varepsilon} \, F_{1,2}(n) \biggr] 
\label{FI1} \\
&& + \biggl[ \ln^2 (-x) \, F_{2,1}(n)  + \frac{1}{\varepsilon} \,\ln (-x) \,
 F_{2,2}(n) + \frac{1}{\varepsilon^2} \, F_{2,3}(n) + \zeta(2)
 \, F_{2,4}(n) \biggr]
+ \cdots \biggr\},
\nonumber
\end{eqnarray}
where $x=q^2/m^2$, $\eta =1$
or $-1$, $c =0$, $1$ and $2$,
and $\alpha=1$ and $2$ for
two-point and three-point cases, respectively.

Here the normalization 
$\hat{N}={(\overline{\mu}^2/m^{2})}^{2\varepsilon}$, 
where $\overline{\mu}=4\pi e^{-\gamma_E} \mu$ is in the standard
$\overline{MS}$-scheme and $\gamma_E$ is the Euler constant.
Moreover, the space-time dimension is $D=4-2\varepsilon$ and
\begin{eqnarray}
C_n  ~=~1, ~~~\mbox{ and }~~~
C_n  ~=~ \frac{(n!)^2}{(2n)!} ~\equiv ~ \hat{C}_n
\label{FI1b}
\end{eqnarray}
for diagrams with
two-massive-particle-cuts ($2m$-cuts). For the diagrams with one-massive-particle-cuts ($m$-cuts)
$C_n = 1$.

For  $m$-cut
case,
the coefficients $F_{N,k}(n)$ should have the form
\begin{eqnarray}
F_{N,k}(n) ~ \sim ~ \frac{S_{\pm a,...}}{n^b}\, ,
\label{FI1c}
\end{eqnarray}
where $S_{\pm a,...} \equiv S_{\pm a,...}(j-1)$
are harmonic sums in (\ref{FI2}).

\begin{figure}[t]
\includegraphics[width=0.6\textwidth,height=0.40\textheight,angle=0]{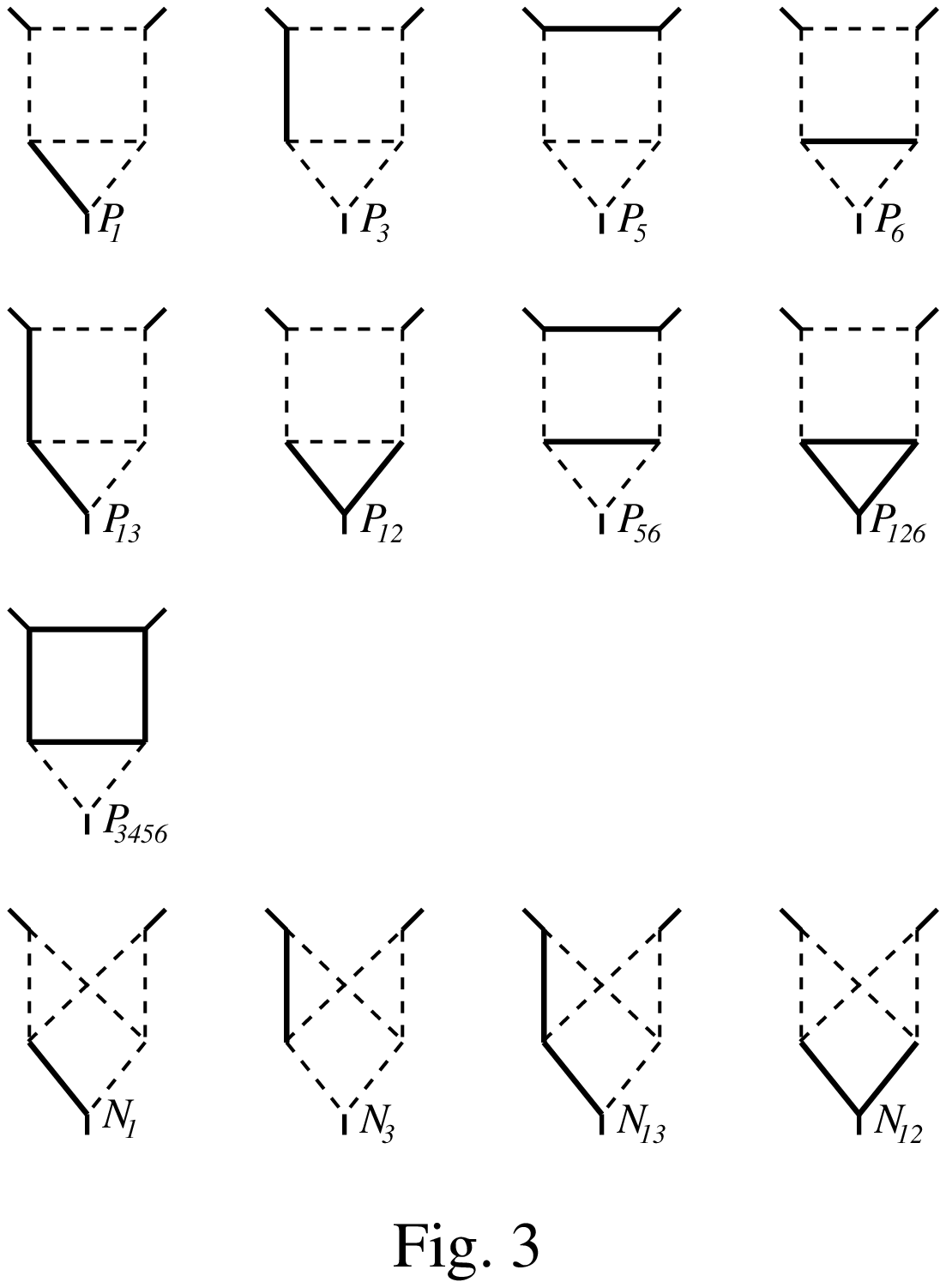}
\end{figure}

For  $2m$-cut
case,
the coefficients $F_{N,k}(n)$ should have the form
\footnote{Really, there are even more complicated terms as ones in Eqs.
(58) and (59) of \cite{FleKoVe}, 
which come from other $\eta $ values in (\ref{FI1}).
However,
they are outside of our present consideration.}
\begin{eqnarray}
F_{N,k}(n) ~ \sim ~ \frac{S_{\pm a,...}}{n^b},  ~ \frac{V_{a,...}}{n^b}
,  ~ \frac{W_{a,...}}{n^b} \, ,
\label{FI1d}
\end{eqnarray}
where $V_{\pm a,...} \equiv V_{\pm a,...}(j-1)$ and 
$W_{\pm a,...} \equiv W_{\pm a,...}(j-1)$ with
\begin{eqnarray}
V_{a}(j)\ =\ \sum^j_{m=1}
\, \frac{\hat{C}_m}{m^a},
\ \ V_{a,b,c,\cdots}(j)~=~ \sum^j_{m=1}  \,
\frac{\hat{C}_m}{m^a}\, S_{b,c,\cdots}(m),  \label{FI4} \\
W_{a}(j)\ =\ \sum^j_{m=1} \,
\frac{\hat{C}_m^{-1}}{m^a},
\ \ W_{a,b,c,\cdots}(j)~=~ \sum^j_{m=1}  \,
\frac{\hat{C}_m^{-1}}{m^a}\, S_{b,c,\cdots}(m),  \label{FI5}
\end{eqnarray}

The terms $\sim V_{a,...}$ and $\sim W_{a,...}$
can come only in the $2m$-cut
case.
%
The origin of the appearance of these  terms 
is the product of series (\ref{FI1})
with the different 
coefficients $C_n =1$ and
$C_n = \hat{C}_n
$.

As an example, consider two-loop two-point diagrams $I_1$
and $I_{12}$ shown in Fig. 2 and  
studied in \cite{FleKoVe}
\begin{eqnarray}
I_1 &=& \frac{\hat{N}}{q^{2}} \,
\sum_{n=1} \, \frac{x^n}{n} \, \biggl\{
\frac{1}{2} \ln^2 (-x) - \frac{2}{n} \ln (-x)  + \zeta(2)
+2S_2 -2 \frac{S_1}{n} + \frac{3}{n^2} \biggr\} \, ,
\label{FI6a} \\
I_{12} &=& \frac{\hat{N}}{q^{2}} \,
\sum_{n=1} \, \frac{x^n}{n^2} \, \biggl\{\frac{1}{n} +
 \frac{(n!)^2}{(2n)!} \, \biggl(
-2 \ln (-x)  -3 W_1 + \frac{2}{n} \biggr) \biggr\} \, .
\label{FI6c}
\end{eqnarray}

From (\ref{FI6a}) 
one can see that the corresponding functions
$F_{N,k}(n)$ have the form 
\begin{eqnarray}
F_{N,k}(n) ~ \sim ~ \frac{1}{n^{2-N}},~~~~(N\geq 2),
\label{FI8}
\end{eqnarray}
if we introduce the following complexity of the sums ($\Phi=(S,V,W)$) 
\begin{eqnarray}
\Phi_{\pm a} \sim \Phi_{\pm a_1, \pm a_2}
\sim \Phi_{\pm a_1,\pm a_{2},\cdots,\pm a_m}
\sim \zeta_{a} \sim \frac{1}{n^a},~~~~ (\sum_{i=1}^m a_i =a) \, .
\label{FI9}
\end{eqnarray}

In Eq. (\ref{FI6c}),
\begin{eqnarray}
F_{N,k}(n) ~ \sim ~ \frac{1}{n^{1-N}},~~~~(N\geq 1),
\label{FI10}
\end{eqnarray}
since now the factor $1/n^2$ has been already extracted.

So, Eqs. (\ref{FI6a})-(\ref{FI6c}) show that the functions $F_{N,k}(n)$
should have the following form
\begin{eqnarray}
\frac{1}{n^{c}} \, F_{N,k}(n) ~ \sim ~ \frac{1}{n^{3-N}},~~~~(N\geq 2)
\label{FI11}
\end{eqnarray}
and the number $3-N$ defines the level of transcendentality (or complexity)
of the coefficients $F_{N,k}(n)$. The property reduces
strongly the number
of the possible elements in $F_{N,k}(n)$.
The level of transcendentality decreases if we consider the
singular parts of diagrams and/or coefficients in front of
$\zeta$-functions and of
logarithm powers.

Other $I$-type integrals in \cite{FleKoVe} have similar form. They have been 
calculated
exactly by differential equation method \cite{Kotikov:1990zs,Kotikov:1990kg}.

Now we consider two-loop three-point diagrams, 
$P_5$
and $P_{12}$ shown in Fig. 3 and calculated
in \cite{FleKoVe}:
\begin{eqnarray}
P_5 &=& \frac{\hat{N}}{(q^{2})^2} \,
\sum_{n=1} \, \frac{(-x)^n}{n} \, \biggl\{
-6\zeta_3 + 2(S_1\zeta_2
+6S_3-2S_1S_2+ 4 \frac{S_2}{n}-\frac{S_1^2}{n}
+ 2\frac{S_1}{n^2} \nonumber \\
&&+ \biggl(-4S_2+S_1^2-2\frac{S_1}{n}\biggr)\ln (-x)
+ S_1\ln^2 (-x)\biggl\} \, ,
\label{FI7b} \\
P_{12} &=& \frac{\hat{N}}{q^{2}} \,
\sum_{n=1} \, \frac{x^n}{n^2} \,
 \frac{(n!)^2}{(2n)!} \, \biggl\{
\frac{2}{\ep^2} + \frac{2}{\ep} \biggl(S_1 -3 W_1 + \frac{1}{n} - \ln (-x)
\biggr) +12 W_2 -18 W_{1,1}
\nonumber \\ && -13S_2 + S_1^2- 6S_1W_1 +2 \frac{S_1}{n} +
\frac{2}{n^2}
-2 \bigg(S_1+\frac{1}{n}\biggr)\ln (-x)  + \ln^2 (-x) \biggr\} \, ,
\nonumber
\end{eqnarray}

 Now the coefficients
$F_{N,k}(n)$ have the form
\begin{eqnarray}
\frac{1}{n^{c}} \, F_{N,k}(n) ~ \sim ~ \frac{1}{n^{4-N}},~~~~(N\geq 3),
\label{FI12}
\end{eqnarray}

The diagram 
$P_5$ 
(and also $P_1$, $P_3$ and $P_6$ in \cite{FleKoVe})
have been calculated
exactly by differential equation method \cite{Kotikov:1990zs,Kotikov:1990kg}.
To find the results for  
$P_{12}$ (and also
all others
in \cite{FleKoVe}) we have used the knowledge of the several $n$ terms in the
inverse-mass expansion (\ref{FI1}) (usually less than $n=100$) and
the following arguments: 
\begin{itemize}
\item
The coefficients should have the structure (\ref{FI12}) with the rule
(\ref{FI9}). The condition (\ref{FI12}) reduces strongly the number of
possible harmonic sums. It should are related with the specific form of the
differential equations for the considered master integrals, like
\bea
\left(\overline{k}\ep  + m^2\frac{d}{dm^2} \right) \, \mbox{ FI } \, = \,
 \mbox{ less complicated diagrams },
\nonumber
\eea
with some $\overline{k}$ values.
We note that for many other master integrals (for example, for sunsets
with two massive lines
in \cite{Kotikov:1990zs,Fleischer:1999hp,Kniehl:2005bc}) the property 
(\ref{FI12}) is violated: the coefficients
$F_{N,k}(n)$ contain sums with different levels of complexity.
\footnote{Really, Refs. \cite{Kotikov:1990zs,Fleischer:1999hp} contain 
the Nilson polylogarithms,
whose sum of indices relates directly to the level of transcendentality
$(4-N)$. The representation of the series (\ref{FI6a})-(\ref{FI6c}) and
 (\ref{FI7b}),
containing $S_{\pm a, \cdots}$, as
polylogarithms can be found in \cite{FleKoVe} for $m$-cut case and in 
\cite{Davydychev:2003mv} for
$2m$-cut one, respectively.}
\item
If
a two-loop two-point diagram with the
``similar topology'' (for example, 
$I_{12}$ for $P_{12}$ an so on)
has been already calculated, we should consider a
similar set of basic elements for corresponding $F_{N,k}(n)$ of
two-loop three-point diagrams
but with the higher level of complexity.
\item
Let the considered diagram contain singularities and/or powers of logarithms.
Because in front of the leading
singularity, or
the largest power of
logarithm, or the largest $\zeta$-function the coefficients are very simple,
they can be often predicted directly from the first several terms of
expansion.

Moreover, often we can calculate the singular part using another technique
(see \cite{FleKoVe} 
for extraction of $\sim W_1(n)$ part). Then we should expand the
singular parts, find the basic elements and try to use them
(with the corresponding increase of the level of complexity) to predict
the regular part of the diagram. If we have to
find the $\ep$-suppressed terms, we should
increase the level of complexity for the corresponding
basic elements.
\end{itemize}

Later, using the ansatz for $F_{N,k}(n)$ and several terms
(usually, less than 100) in the above
expression, which can be calculated exactly, we obtain the system of
algebraic equations for the parameters of the ansatz. Solving the system, we
can obtain the analytical results for FI without
exact calculations.
To check the results, it is needed only to calculate a few more
terms
in the
above inverse-mass expansion (\ref{FI1}) and compare them with the
predictions of our anzatz with the above fixed coefficients.

So, the considered arguments give a possibility to find the results for many complicated
two-loop three-point diagrams without direct calculations.
Some variations of the procedure have been successfully used for calculating
the Feynman diagrams for many processes (see 
\cite{Fleischer:1997bw,FleKoVe,Kniehl:2005bc,Kniehl:2006bg}).

Nore that the properties similar to (\ref{FI11}) and (\ref{FI12}) have been
observed recently \cite{Eden:2012rr} in the so-called double
operator-product-expansion limit of some four-point diagrams. These diagrams
are encoded the quantum corrections to the four-point correlator and
have been considered in \cite{Eden:2012rr} up-to three-loop level of accuracy.

\section{Conclusion}
In the first part of this short review we presented the universal AD
$\gamma _{uni}(j)$ for
the ${\mathcal N}=4$ supersymmetric gauge theory up to the NNLO
approximation. All the results have been obtained with using of the
{\it transcendentality principle}.
At the first three orders, the universal anomalous dimension
have been extracted from the corresponding QCD calculations.
The results for four and five loops have been obtained from
the long-range asymptotic Bethe equations
together with some additional terms, so-called {\it wrapping
corrections}, coming in agreement with Luscher approach.

The second part contains the consideration of so-called master integrals, which obey also to
the similar {\it transcendentality principle} (\ref{FI9}).
Its application leads to the possibility to get the results for most of master integrals
without direct calculations.

This work was supported by 
RFBR grant 10-02-01259-a.
Author 
thanks the Organizing Committee of IV  International Conference ``Models in Quantum Field Theory''
(MQFT-2012)
for invitation.

\end{document}